\documentclass[conference]{IEEEtran}
\IEEEoverridecommandlockouts

\usepackage{cite}
\usepackage{stmaryrd}
\newcommand{\enc}[1]{\llbracket #1 \rrbracket}
\usepackage{amsmath,amssymb,amsfonts}
\usepackage{algorithm}
\usepackage{algorithmic}
\usepackage{graphicx}
\usepackage{booktabs}
\usepackage{multirow}

\usepackage{textcomp}
\usepackage{xcolor}
\usepackage{url}
\usepackage{marvosym}
\usepackage{balance}
\usepackage{dblfloatfix}
\usepackage{placeins}

\def\BibTeX{{\rm B\kern-.05em{\sc i\kern-.025em b}\kern-.08em
T\kern-.1667em\lower.7ex\hbox{E}\kern-.125emX}}

\title{TGHE: Template-based Graph Homomorphic Encryption for Privacy-Preserving GNN Inference in Edge-Cloud Systems}

\author{
\begin{tabular}{@{}c@{\hspace{0.35in}}c@{\hspace{0.35in}}c@{}}
\begin{tabular}[t]{c}
Ngoc Bao Anh Le\textsuperscript{\Letter}\\
\textit{University of Wollongong}\\
Wollongong, Australia\\
nbal745@uowmail.edu.au\\
0009-0005-5373-3149
\end{tabular}
&
\begin{tabular}[t]{c}
Thai T. Vu\textsuperscript{\Letter}\\
\textit{University of Wollongong}\\
Wollongong, Australia\\
tienv@uow.edu.au\\
0000-0002-9826-3321
\end{tabular}
&
\begin{tabular}[t]{c}
John Le\\
\textit{University of Wollongong}\\
Wollongong, Australia\\
johnle@uow.edu.au\\
0000-0003-0019-0345
\end{tabular}
\end{tabular}\\[1.5ex]
\begin{tabular}{@{}c@{\hspace{0.6in}}c@{}}
\begin{tabular}[t]{c}
Heath Cooper\\
\textit{University of Wollongong}\\
Wollongong, Australia\\
hcooper@uow.edu.au
\end{tabular}
&
\begin{tabular}[t]{c}
Jun Shen\\
\textit{University of Wollongong}\\
Wollongong, Australia\\
jshen@uow.edu.au\\
0000-0002-9403-7140
\end{tabular}
\end{tabular}
}

\begin{document}
\maketitle

\begin{abstract}
Existing homomorphic encryption (HE)-based GNN systems adopt a graph-centric paradigm that couples per-query cost to global graph size, limiting evaluations to at most $\sim$20k nodes and making them incompatible with dynamic, large-scale financial graphs. We propose TGHE (Template-based Graph Homomorphic Encryption), an ego-centric framework that resolves this by exploiting a \emph{template phenomenon}: local computation trees in transaction graphs converge into a small set of structural shapes. TGHE canonicalizes ego-graphs at the edge and packs structurally identical trees into shared CKKS ciphertexts for SIMD-parallel encrypted inference, with two long-tail optimizers (\emph{Approximate Template Fitting} and \emph{Topology Collapse}) ensuring full SIMD coverage. On DGraphFin (3.7M nodes, 4.3M edges), TGHE-Collapse achieves a 66.9$\times$ speedup over the sequential encrypted baseline with less than 0.002 AUC loss.
\end{abstract}


\section{Introduction}
\label{sec:intro}

Financial institutions increasingly delegate analytics workloads to cloud platforms for scalability, yet regulatory and trust constraints prohibit exposing raw transaction data to third-party infrastructure. This tension is acute in fraud detection, which increasingly relies on Graph Neural Networks (GNNs)~\cite{corso2024graph} to capture relational patterns among accounts and transactions. In production systems, both node features (e.g., transaction statistics) and graph topology (e.g., interaction patterns) are highly sensitive. Protecting this data during inference is critical, especially in cross-institution or cloud-assisted deployments where raw data must not leave the trusted perimeter.

Homomorphic Encryption (HE)~\cite{acar2018survey} enables computation on encrypted data without decryption, making it a strong candidate for privacy-preserving inference. Recent HE-based GNN systems, including CryptoGCN~\cite{ran2022cryptogcn}, LinGCN~\cite{peng2023lingcn}, Penguin~\cite{ran2023penguin}, and FicGCN~\cite{ficgcn2024}, have demonstrated feasibility on standard benchmarks. However, these systems share a \emph{transductive, graph-centric} design: they require encrypting and operating over global adjacency matrices. This creates three practical limitations. First, encrypted matrix operations scale poorly; existing evaluations are limited to graphs with at most tens of thousands of nodes. Second, a static global graph is assumed, which is incompatible with dynamic financial systems where new accounts appear continuously. Third, structural preprocessing such as partitioning or reordering may itself leak information or add deployment complexity.

Real-world fraud detection operates differently. Transaction data originates at distributed edge locations (e.g., bank branches), and low-latency decisions are required locally. This setting naturally favors \emph{ego-centric} inference: classifying a queried node based on its local neighborhood rather than the entire graph. The resulting query-response pattern, where each edge node submits an encrypted local subgraph and receives an encrypted prediction, aligns with on-demand cloud-assisted deployment.

In this paper, we propose \textbf{TGHE} (Template-based Graph Homomorphic Encryption), a framework for efficient encrypted GNN inference in edge-cloud systems. TGHE is built on an empirical insight we call the \emph{template phenomenon}: despite the scale and irregularity of financial graphs, local computation trees converge into a small set of structural shapes. We validate this across four transaction graph datasets. TGHE exploits this regularity through canonicalization, template-based ciphertext packing, and two optimizers that maximize SIMD coverage.

We evaluate the full TGHE pipeline on DGraphFin~\cite{pyg2025dgraphfin,huang2022dgraph}, a financial graph with 3.7 million nodes and 4.3 million edges - orders of magnitude larger than prior HE-GNN evaluations. The contributions are:

\begin{itemize}
    \item \textbf{Ego-centric encrypted inference.} We decouple query-time computation from the global graph by extracting and canonicalizing local ego-graphs at the edge. Per-query cost is independent of graph size, enabling deployment in dynamic edge-cloud environments.

    \item \textbf{Template-based SIMD packing.} We design a canonicalization and template-matching strategy that groups structurally identical ego-graphs into shared CKKS ciphertexts~\cite{cheon2017ckks} for uniform SIMD-parallel evaluation. We validate the underlying template phenomenon across four financial datasets of varying scale and density.

    \item \textbf{Long-tail optimizers.} We propose Approximate Template Fitting, which maps unmatched ego-graphs to nearby templates via mean-preserving padding, and Topology Collapse, which pre-aggregates hop-2 neighborhoods in plaintext to eliminate structural diversity entirely. TGHE-Collapse achieves a 66.9$\times$ end-to-end speedup over the sequential encrypted baseline with less than 0.002 AUC loss.
\end{itemize}
\section{Related Work}
\label{sec:related_work}

Existing HE-based GNN systems target GCN-style models under the CKKS scheme. CryptoGCN~\cite{ran2022cryptogcn} exploits adjacency sparsity to reduce ciphertext operations. LinGCN~\cite{peng2023lingcn} lowers multiplicative depth through linearization. Penguin~\cite{ran2023penguin} reduces rotation count via two-dimensional ciphertext packing. FicGCN~\cite{ficgcn2024} addresses irregular graph structures with latency-aware packing. These systems share a common \emph{graph-centric} execution model: inference requires encrypting and operating over complete or partitioned adjacency matrices, coupling per-inference cost to the input graph size and limiting evaluations to at most $\sim$20k nodes (Table~\ref{tab:related_comparison}). TGHE adopts ego-centric inference, decoupling per-query cost from graph size and enabling evaluation on DGraphFin~\cite{pyg2025dgraphfin,huang2022dgraph} (3.7M nodes), orders of magnitude beyond prior work. Because prior systems' throughput grows with global graph size while TGHE's per-query cost is graph-size-independent, direct runtime comparison across the two paradigms is not meaningful; Table~\ref{tab:related_comparison} therefore reports evaluation scale rather than latency. GNN topology leakage has been studied from an attack perspective~\cite{he2021stealing,fu2025safeguarding}; we address it at the system level through a leakage-aware design characterized in Section~\ref{sec:exp_leakage}.

\begin{table}[!htbp]
\centering
\caption{Comparison of HE-based GNN inference frameworks.}
\label{tab:related_comparison}
\small
\begin{tabular}{lcccc}
\toprule
 & \textbf{Task} & \textbf{Preprocessing} & \textbf{Eval.\ Scale} \\
\midrule
CryptoGCN & Graph cls. & Sparsity formatting & $\sim$25 \\
LinGCN & Graph cls. & Linearization & $\sim$25 \\
Penguin & Node cls. & 2D packing & $\sim$20k \\
FicGCN & Node cls.\textsuperscript{$\ddagger$} & Reordering & $\sim$20k \\
\midrule
\textbf{TGHE} & \textbf{Node cls.} & \textbf{Client-side canon.} & \textbf{3.7M\textsuperscript{$\dagger$}} \\
\bottomrule
\end{tabular}
\vspace{2pt}
\par\noindent\footnotesize \emph{Eval.\ scale}: largest evaluated graph. All prior methods require global adjacency structures. \textsuperscript{$\dagger$}Per-query cost is independent of global graph size. \textsuperscript{$\ddagger$}FicGCN also evaluates graph-level classification on skeleton graphs ($\sim$25 nodes).
\end{table}

\section{System Model and Problem}
\label{sec:system_model}

\begin{figure}[!htbp]
  \centering
  \includegraphics[width=\linewidth]{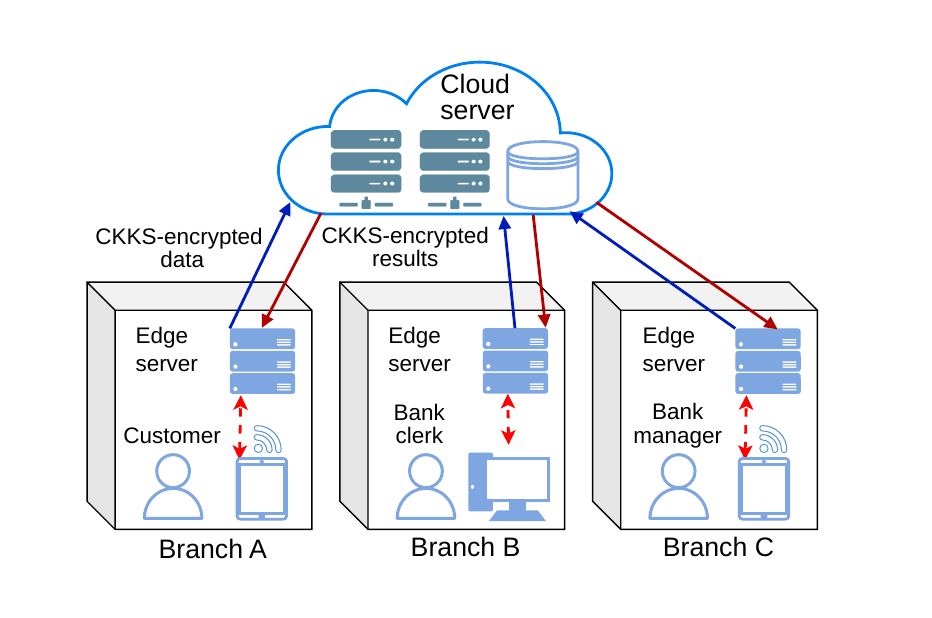}
  \caption{Edge-cloud architecture for encrypted GNN inference. Edge servers extract, canonicalize, and encrypt local ego-graphs; the cloud performs inference on ciphertexts.}
  \label{fig:edge_cloud_model}
\end{figure}

We model financial transaction data as a directed graph $\mathcal{G} = (\mathcal{V}, \mathcal{E})$ where nodes represent entities (e.g., accounts) and edges represent transactions, each node $v$ having a feature vector $\mathbf{x}_v$. As shown in Fig.~\ref{fig:edge_cloud_model}, edge servers are co-located with data sources and hold local transaction records. For each query node $v$, the edge extracts a local $L$-hop ego-graph $\mathcal{G}_v$, applies TGHE canonicalization, and encrypts it under CKKS. The cloud holds trained parameters $\Theta$, performs inference entirely on ciphertexts, and returns encrypted logits $\enc{\hat{y}_v} = f_\Theta(\enc{\mathcal{G}_v})$ to the edge for decryption, without observing raw features or outputs.

\smallskip
\noindent\textbf{Threat model.} The cloud is \emph{honest-but-curious}; the edge client is trusted and holds the CKKS secret key. Node features and model outputs are protected by CKKS semantic security~\cite{cheon2017ckks} under RLWE hardness. We adopt a \emph{leakage-aware} design: SIMD batching requires the server to receive coarse structural metadata (template signatures, degree counts), and edge attributes are transmitted in plaintext for correction terms; all residual leakage is explicitly characterized in Section~\ref{sec:exp_leakage}.

\smallskip
\noindent\textbf{CKKS and GNN notation.} We denote encryption by $\enc{\cdot}$. CKKS supports SIMD computation over packed ciphertext vectors; dominant costs are slot rotations and ciphertext-ciphertext (CT-CT) multiplications, while ciphertext-plaintext (CT-PT) operations are orders of magnitude cheaper. A standard GNN layer computes:
\begin{equation}
\mathbf{H}^{(l+1)} = \sigma\!\left(\tilde{\mathbf{A}}\,\mathbf{H}^{(l)}\mathbf{W}^{(l)}\right),
\label{eq:gnn_layer}
\end{equation}
where $\tilde{\mathbf{A}}$ is the normalized adjacency, $\mathbf{W}^{(l)}$ are learnable parameters, and $\sigma(\cdot)$ is approximated by a low-degree polynomial under HE.

\smallskip
\noindent\textbf{Problem.} Given $\mathcal{G}$ with sensitive features and topology, compute GNN predictions for each query node $v \in \mathcal{Q}$ using only encrypted local neighborhoods. The central challenge is a \emph{structural mismatch}: HE requires uniform arithmetic circuits, but ego-graphs vary in degree, depth, and connectivity. Section~\ref{sec:methodology} resolves this mismatch.

\section{Proposed Method}
\label{sec:methodology}

As established in Section~\ref{sec:system_model}, HE requires uniform arithmetic circuits, but ego-graphs vary in structure. TGHE resolves this mismatch through a three-stage pipeline:

\begin{enumerate}
  \item The edge client extracts and canonicalizes each ego-graph into a deterministic computation tree (Section~\ref{sec:ego_centric}).
  \item Structurally identical trees are grouped by \emph{template signature} and packed into shared CKKS ciphertexts for SIMD-parallel evaluation (Section~\ref{sec:template_packing}).
  \item Unmatched trees are handled by two long-tail optimizers that maximize SIMD coverage (Section~\ref{sec:long_tail}).
\end{enumerate}

\noindent Stages 1-2 are architecture-agnostic. Stage 3 includes one optimizer (Topology Collapse) that exploits the branch-separated structure of the target GNN; the architecture is specified in Section~\ref{sec:topo_collapse} where this property becomes essential.

\subsection{Ego-Graph Extraction and Canonicalization}
\label{sec:ego_centric}

For each query node $v \in \mathcal{Q}$, the edge client extracts a 2-hop ego-graph from the local transaction data, matching the receptive field of the target two-layer GNN. This subgraph is irregular: nodes have varying degrees, shared neighbors create cycles, and multi-paths violate SIMD uniformity. The client transforms it into a canonical computation tree through two plaintext steps.

\subsubsection{Deterministic Capping}
The client caps each ego-graph to bounded size: at most $d_{\max}$ hop-1 neighbors per root, and $c_{\max}$ hop-2 neighbors per hop-1 node ($d_{\max}{=}10$, $c_{\max}{=}5$ in experiments). Hop-1 neighbors are selected by degree (highest retained); hop-2 neighbors are selected by node index. Capping bounds the computation tree size, which in turn bounds the HE circuit depth and ciphertext count.

\subsubsection{Treeification}
\begin{figure}[!htbp]
  \centering
  \includegraphics[width=\linewidth]{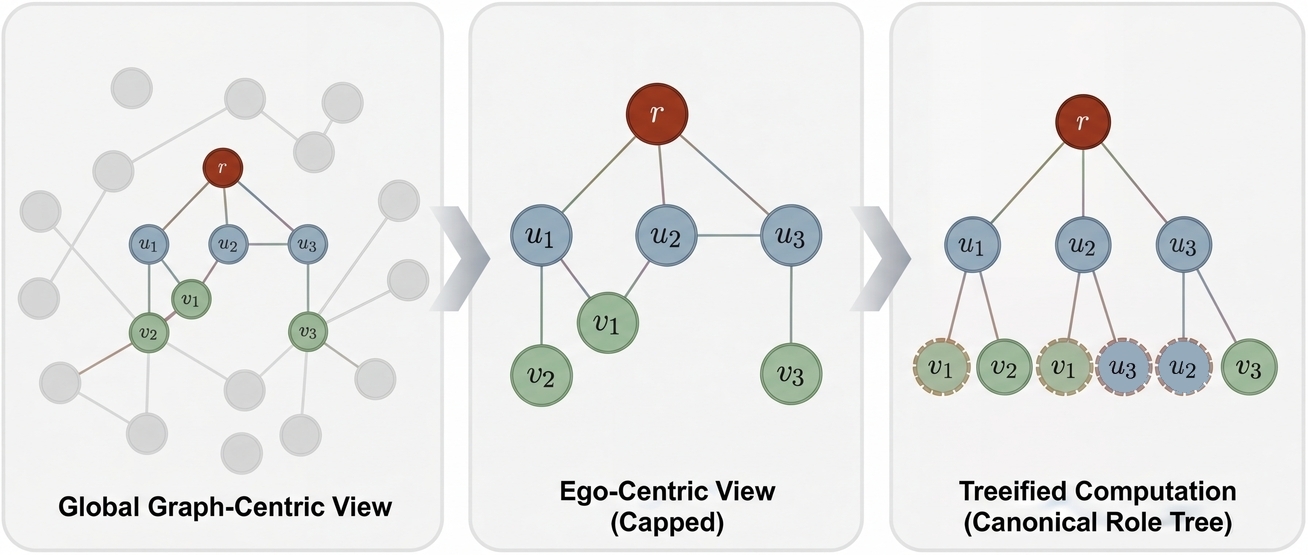}
  \caption{From global graph to canonical computation tree. Left: the node's neighborhood in the global graph. Center: the extracted ego-graph after capping. Right: the treeified role tree with cloned nodes.}
  \label{fig:global_to_ego_to_tree}
\end{figure}

Even after capping, the subgraph may contain shared nodes and cycles. The client resolves these by \emph{treeification}: if a hop-2 node $w$ is reachable from multiple hop-1 parents, it is cloned independently under each parent, retaining the original feature $\mathbf{x}_w$. The result is a deterministic role tree where each position maps to exactly one feature vector. Fig.~\ref{fig:global_to_ego_to_tree} illustrates the full transformation from global graph to canonical tree.

\subsection{The Template Phenomenon and SIMD Packing}
\label{sec:template_packing}

After canonicalization, each ego-graph is a bounded tree whose depth (two hops) and breadth (degree capping) limit the space of possible shapes. The key question is: how many \emph{distinct} tree shapes exist across all queries? If the answer is small, we can batch structurally identical trees into shared ciphertexts for SIMD-parallel evaluation.

\subsubsection{Template Signatures}
We capture each tree's shape by its \emph{template signature}, where $r$ denotes the root of the canonicalized tree:
\begin{equation}
  \mathrm{sig}(r) = \bigl(d,\; (c_1, c_2, \ldots, c_d)\bigr), \quad c_1 \geq c_2 \geq \cdots \geq c_d,
  \label{eq:template_sig}
\end{equation}
where $d$ is the hop-1 degree after capping and $c_j$ is the number of hop-2 children under the $j$-th hop-1 neighbor, sorted in descending order. Two trees with the same signature have identical role structure and require the same HE circuit.

\subsubsection{Empirical Observation}
\label{sec:template_observation}

\begin{figure}[!htbp]
  \centering
  \includegraphics[width=\columnwidth]{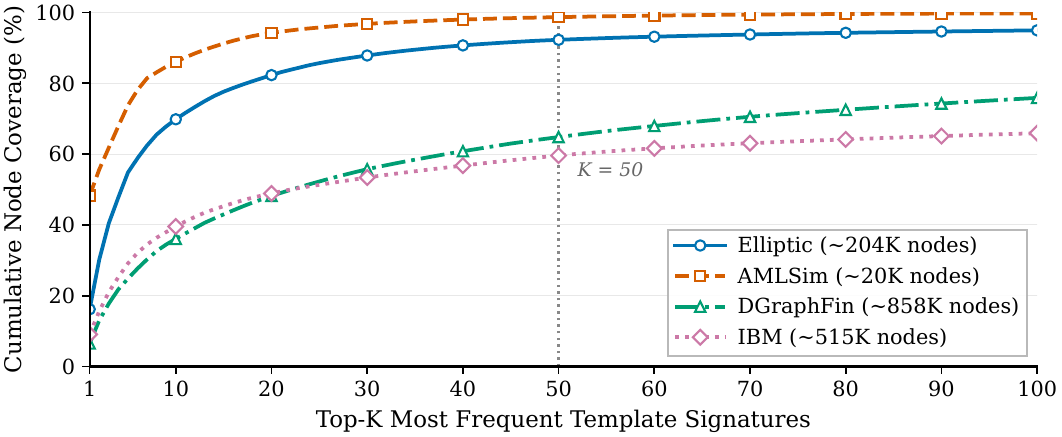}
  \caption{Cumulative node coverage vs.\ template library size ($K$) across four financial transaction datasets. All datasets exhibit rapid convergence, with coverage varying by graph density and degree heterogeneity.}
  \label{fig:template_coverage}
\end{figure}

Despite the massive irregularity of real-world graphs, the number of distinct template signatures is far smaller than the number of query nodes. Fig.~\ref{fig:template_coverage} shows the cumulative query coverage as a function of template library size $K$ over the query subsets used for template analysis on four financial transaction datasets. Two patterns emerge. First, all datasets exhibit rapid initial convergence: a small number of templates covers a disproportionate fraction of queries. Second, the coverage level varies with graph characteristics: denser, more homogeneous graphs (Elliptic~\cite{weber2019elliptic}, AMLSim~\cite{suzumura2021amlsim}) reach over 90\% coverage at $K{=}50$, while larger, more heterogeneous graphs (DGraphFin~\cite{pyg2025dgraphfin,huang2022dgraph} and IBM~\cite{altman2023ibmaml}) reach approximately 60\% when evaluated on sampled query subsets.

This \emph{template phenomenon} arises from degree capping: once the neighborhood is bounded to $d_{\max}$ hop-1 and $c_{\max}$ hop-2 neighbors, the combinatorial space of possible tree shapes is finite. In practice, degree distributions in transaction graphs are concentrated enough that a small subset of shapes dominates.

\subsubsection{Template-Batched SIMD Packing}
Queries sharing a template signature have identical role trees. The client groups up to $C$ such queries and packs their feature vectors into shared CKKS ciphertexts, one ciphertext per tree role. A batch of $C$ queries with signature $(d, (c_1, \ldots, c_d))$ requires:
\begin{equation}
  N_{\mathrm{CT}} = 1 + d + \textstyle\sum_{j=1}^{d} c_j
  \label{eq:nct}
\end{equation}
ciphertexts (one for the root, one per hop-1 node, one per hop-2 node). Features are arranged in a \emph{duplicated layout} with block size $b$ for efficient DiagBSGS (Baby-Step Giant-Step rotation algorithm for matrix-vector multiplication)\cite{halevi2018faster} matrix-vector multiplication:
\begin{equation}
  C = \left\lfloor \frac{N/2}{2b} \right\rfloor.
  \label{eq:batch_capacity}
\end{equation}

With CKKS polynomial degree $N{=}16{,}384$, each ciphertext provides $N/2{=}8{,}192$ plaintext slots. With block size $b{=}32$, each ciphertext holds $C{=}128$ queries. Queries sharing a template are grouped and their feature vectors packed into shared ciphertexts (one per tree role), so the server executes a single HE circuit across all SIMD slots, amortizing the rotation schedule over 128 queries simultaneously.

\subsubsection{The Long-Tail Bottleneck}
\label{sec:longtail_bottleneck}

Template-based packing handles the majority of queries, but the remaining unmatched queries fall back to sequential, per-query encrypted execution. This fallback is disproportionately expensive. Each fallback query independently executes the full DiagBSGS rotation schedule, consuming the same number of rotations as an entire packed batch of 128 queries. On DGraphFin with $K{=}50$ templates, TGHE-Base routes 99\% of queries through the packed path, yet the 1\% fallback still accounts for over 70\% of total rotations (Table~\ref{tab:e2e}). This is a direct instance of Amdahl's Law: even a small sequential fraction bounds overall speedup.

\subsection{Long-Tail Optimizers}
\label{sec:long_tail}

We propose two complementary strategies to eliminate the long-tail bottleneck. \emph{Approximate Template Fitting} (Section~\ref{sec:approx_fitting}) is architecture-agnostic and maps unmatched ego-graphs to nearby templates via padding. \emph{Topology Collapse} (Section~\ref{sec:topo_collapse}) exploits properties of the target GNN architecture to eliminate hop-2 structural diversity entirely, collapsing all templates into degree groups with zero fallback.

\subsubsection{Approximate Template Fitting}
\label{sec:approx_fitting}

If a query's tree does not exactly match any template, the client searches for the closest template and pads the tree to fit. The padding must preserve aggregation semantics so that the encrypted output remains correct.

For each unmatched root $r$, the client searches the template library $\mathcal{T}$ for the best approximate match. For each candidate template $T$ with hop-1 degree $d_T \leq d_{\mathrm{rt}}$, the client selects a subset $S$ of $d_T$ hop-1 roles, assigns real hop-2 nodes, and fills excess slots with \emph{surrogate features}. Surrogates are set to the mean of real neighbors in that role's aggregation group (root plus real hop-2 children):
\begin{equation}
  \bar{\mathbf{x}} = \mathrm{mean}(\mathbf{x}_r, \mathbf{x}_{w_1}, \ldots, \mathbf{x}_{w_m}).
  \label{eq:surrogate}
\end{equation}
This choice preserves the neighbor-branch mean exactly: if a role has $m$ real neighbors and the template requires $c > m$ slots, padding with $\bar{\mathbf{x}}$ leaves the aggregated mean unchanged.

Match quality is measured by:
\begin{equation}
  \mathrm{sim}(T, S) = \underbrace{\frac{1 + d_T + n_{\text{real-hop2}}}{1 + d_T + \sum_{j=1}^{d_T} c_j}}_{\text{fill fraction}} \times \underbrace{\frac{d_T}{d_{\mathrm{rt}}}}_{\text{keep ratio}}.
  \label{eq:sim_score}
\end{equation}
The fill fraction measures how many template slots are occupied by real data rather than surrogates. The keep ratio penalizes dropping hop-1 neighbors. If the best score exceeds a threshold $\tau$, the query is fitted to that template; otherwise it falls back (Algorithm~\ref{alg:approx_fitting}). The search space $|\mathcal{T}| \times \binom{d_{\mathrm{rt}}}{d_T}$ is tractable under capping ($d_{\mathrm{rt}} \leq 10$). Surrogate features are not masked during aggregation, avoiding per-slot metadata leakage. Empirically, AUC degradation is below 0.003 at $\tau{=}0.50$.

\begin{algorithm}[t]
\caption{Approximate Template Fitting}
\label{alg:approx_fitting}
\begin{algorithmic}[1]
\REQUIRE Root $r$; library $\mathcal{T}$; caps $(d_{\max}, c_{\max})$; threshold $\tau$
\ENSURE Packed-path match or fallback
\STATE Treeify, cap, compute $\mathrm{sig}(r)$.
\IF{$\mathrm{sig}(r) \in \mathcal{T}$}
  \RETURN \textsc{ExactMatch}$(\mathrm{sig}(r))$
\ENDIF
\STATE $\mathrm{best\_sim} \leftarrow 0$, $\mathrm{best\_T} \leftarrow \emptyset$, $\mathrm{best\_S} \leftarrow \emptyset$
\FOR{each $T \in \mathcal{T}$ with $d_T \leq d_{\mathrm{rt}}$}
  \FOR{each subset $S$ of hop-1 roles, $|S| = d_T$}
    \STATE Order $S$; assign hop-2; pad surrogates; compute $\mathrm{sim}(T, S)$.
    \IF{$\mathrm{sim}(T, S) > \mathrm{best\_sim}$}
      \STATE Update $\mathrm{best\_sim}$, $\mathrm{best\_T}$, $\mathrm{best\_S}$.
    \ENDIF
  \ENDFOR
\ENDFOR
\IF{$\mathrm{best\_sim} \geq \tau$}
  \RETURN \textsc{ApproxMatch}$(\mathrm{best\_T}, \mathrm{best\_S})$
\ELSE
  \RETURN \textsc{Fallback}
\ENDIF
\end{algorithmic}
\end{algorithm}

\subsubsection{Topology Collapse}
\label{sec:topo_collapse}

Approximate Template Fitting reduces fallback but cannot eliminate it entirely. A stronger strategy is possible when the GNN architecture has a specific property: \emph{linear neighbor aggregation with explicit branch separation}. We now describe the target architecture and the resulting optimization.

\paragraph{Target architecture}
TGHE targets a GraphSAGE variant with explicit self/neighbor branch separation and learned edge-type embeddings, referred to as GEARSage following the implementation by~\cite{gearsage2022}, which holds the top position on the DGraphFin leaderboard (AUC 0.8460). Each layer separates computation into a self branch and a neighbor branch:
\begin{equation}
  \mathbf{h}_v^{(l+1)} = \underbrace{\mathbf{W}_{\mathrm{self}}^{(l)}\,\mathbf{h}_v^{(l)}}_{\text{self}} + \underbrace{\mathbf{W}_{\mathrm{neigh}}^{(l)}\,\mathrm{AGG}\!\left(\{\mathbf{m}_{u \to v}\}\right)}_{\text{neighbor}} + \underbrace{\mathbf{e}_v^{(l)}}_{\text{edge/time}}.
  \label{eq:gearsage_layer}
\end{equation}
Neighbor messages $\mathbf{m}_{u \to v}$ concatenate neighbor features with learned edge-type, direction, and time embeddings before mean-pool aggregation. Branches are fused by summation; the edge/time correction term is computed from plaintext metadata (Section~\ref{sec:exp_leakage}). A trainable polynomial $f(x) = ax^2 + bx + c$ replaces ReLU between layers for HE compatibility.

Two properties make this architecture amenable to Topology Collapse. First, branch separation means the self branch $\mathbf{W}_{\mathrm{self}}^{(l)}\mathbf{h}_v^{(l)}$ depends only on the node itself, with no hop-2 involvement. Second, the neighbor-branch aggregation is linear and commutes with $\mathbf{W}_{\mathrm{neigh}}^{(l)}$, so aggregation can be performed \emph{before} encryption.

\paragraph{Pre-aggregation}
Consider the Layer-1 neighbor-branch aggregation for a hop-1 node $u_j$. In the ego-centric formulation, the root $r$ is connected to each hop-1 node via an outgoing edge and acts as one of its neighbors. The aggregation is:
\begin{equation}
  \bar{\mathbf{n}}_{u_j} = \frac{1}{1{+}|C_j|}\!\left(\mathbf{x}_r + \textstyle\sum_{w \in C_j} \mathbf{x}_w\right),
  \label{eq:gearsage_hop2_agg}
\end{equation}
where $C_j$ denotes the hop-2 children of $u_j$. Because this sum is linear, the client can compute $\mathbf{s}_{u_j} = \mathbf{x}_r + \sum_{w \in C_j} \mathbf{x}_w$ in plaintext and encrypt only the result $\enc{\mathbf{s}_{u_j}}$ alongside $\enc{\mathbf{x}_{u_j}}$. The server then applies the scalar $\frac{1}{1+|C_j|}$ as a cheap CT-PT multiplication (receiving $|C_j|$ as metadata), followed by $\mathbf{W}_{\mathrm{neigh}}^{(0)}$, and adds the self branch and edge/time correction.

\begin{algorithm}[t]
\caption{Topology Collapse}
\label{alg:topo_collapse}
\begin{algorithmic}[1]
\REQUIRE Root $r$; caps $(d_{\max}, c_{\max})$; degree groups $\{G_1,\ldots,G_{d_{\max}}\}$
\ENSURE Exact packed-path execution with no fallback
\STATE Treeify and cap the ego-graph of $r$.
\STATE Order hop-1 roles; compute effective degree $d_{\mathrm{eff}}(r)$.
\FOR{each hop-1 node $u_j$}
  \STATE Collect its hop-2 children $C_j$.
  \STATE Pre-aggregate in plaintext: $\mathbf{s}_{u_j} \leftarrow \mathbf{x}_r + \sum_{w \in C_j}\mathbf{x}_w$.
  \STATE Encrypt $\mathbf{x}_{u_j}$ and $\mathbf{s}_{u_j}$; record $|C_j|$ as metadata.
\ENDFOR
\STATE Pack the query into degree group $G_{d_{\mathrm{eff}}(r)}$.
\STATE Apply mean normalization from recorded child counts.
\STATE Evaluate Layer 1 in packed form using self/neighbor branches and edge/time correction.
\STATE Continue Layer 2 with the exact packed kernel for $G_{d_{\mathrm{eff}}(r)}$.
\RETURN \textsc{ExactDegreeGroup}$(d_{\mathrm{eff}}(r))$
\end{algorithmic}
\end{algorithm}

\paragraph{Structural consequences}
Pre-aggregation eliminates all hop-2 nodes from the encrypted computation. The template signature $(d, (c_1, \ldots, c_d))$ collapses to a single integer: the effective hop-1 degree $d_{\mathrm{eff}}(r)$. Ciphertexts per batch drop from $1 + d + \sum c_j$ to $1 + 2d_{\mathrm{eff}}$ (one root, one self-branch and one neighbor-branch ciphertext per hop-1 node). The template library reduces to at most $d_{\max}$ degree groups, and every query matches exactly. The fallback path is eliminated entirely, yielding a 66.9$\times$ speedup over sequential execution (Section~\ref{sec:exp_evaluation}). The full procedure is summarized in Algorithm~\ref{alg:topo_collapse}.

\paragraph{Applicability}
This equivalence holds for encrypted feature computation in Layer 1. Layer-2 aggregation cannot be collapsed because of the intervening polynomial nonlinearity. Edge/time correction terms are handled separately (Section~\ref{sec:exp_leakage}). Topology Collapse exploits the linearity of mean-pool aggregation, which is the dominant paradigm for high-performing financial GNN models including the leaderboard-leading GEARSage. For attention-based architectures such as GAT, Approximate Template Fitting remains applicable as the general-purpose optimizer.
\section{Experimental Evaluation}
\label{sec:exp_evaluation}

We evaluate TGHE on end-to-end encrypted inference, addressing three questions: (1)~Does encryption degrade predictive accuracy? (2)~How much speedup does template-based packing provide? (3)~What information does the server observe?

\subsection{Experimental Setup}
\label{sec:exp_setup}

All experiments run on a Google Cloud \texttt{n2-standard-16} VM (16 vCPUs, 64\,GB RAM). We use Microsoft SEAL~\cite{sealcrypto2024} with CKKS at polynomial degree $N{=}16{,}384$ (128-bit security, $N/2{=}8{,}192$ slots per ciphertext, batch capacity $C{=}128$). The dataset is DGraphFin~\cite{pyg2025dgraphfin,huang2022dgraph} (3.7M nodes, 4.3M edges). We deploy a two-layer GEARSage~\cite{gearsage2022} with $D_{\mathrm{in}}{=}17$, $D_{\mathrm{hidden}}{=}32$, and trainable quadratic polynomial activations ($d_{\max}{=}10$, $c_{\max}{=}5$). The model is trained transductively in plaintext; TGHE operates exclusively at inference time over encrypted local ego-graphs. We evaluate 10{,}000 randomly sampled query nodes under three modes: \textbf{Non-SIMD} (sequential baseline), \textbf{TGHE-Base} ($K{=}50$, $\tau{=}0.50$), and \textbf{TGHE-Collapse} (Topology Collapse, 100\% SIMD coverage).

\begin{table}[t]
\centering
\caption{Test ROC-AUC on DGraphFin.}
\label{tab:utility}
\small
\begin{tabular}{lc}
\toprule
\textbf{Method} & \textbf{Test AUC} \\
\midrule
\multicolumn{2}{l}{\textit{Encrypted inference}\textsuperscript{$\star$}} \\[2pt]
Non-SIMD (Baseline) & 0.7973 \\
\textbf{TGHE-Collapse} & \textbf{0.7959} \\
\textbf{TGHE-Base} ($\tau{=}0.50$) & \textbf{0.7950} \\
\midrule
\multicolumn{2}{l}{\textit{Plaintext reference}\textsuperscript{$\dagger$}} \\[2pt]
GEARSage~\cite{gearsage2022} & 0.8460 \\
Graph Transformer~\cite{yun2019graph_transformer} & 0.7838 \\
SIGN~\cite{frasca2020sign} & 0.7823 \\
GraphSAGE~\cite{hamilton2017graphsage} & 0.7761 \\
GAT~\cite{velickovic2018gat} & 0.7333 \\
GCN~\cite{kipf2017gcn} & 0.7078 \\
\bottomrule
\end{tabular}
\vspace{2pt}
\par\noindent\footnotesize \textsuperscript{$\star$}HE-adapted GEARSage with polynomial activations and degree capping. \textsuperscript{$\dagger$}Standard architectures with ReLU and full neighborhoods; direct numerical comparison is not meaningful.
\end{table}

\begin{table*}[!t]
\centering
\caption{End-to-end inference performance over 10{,}000 query nodes.}
\label{tab:e2e}
\small
\begin{tabular}{lccccccccccc}
\toprule
& & \multicolumn{3}{c}{\textbf{Routing}} & \multicolumn{2}{c}{\textbf{Rotations}} & \multicolumn{2}{c}{\textbf{P.\ Operations}} & \multicolumn{2}{c}{\textbf{F.\ Operations}} & \\
\cmidrule(lr){3-5} \cmidrule(lr){6-7} \cmidrule(lr){8-9} \cmidrule(lr){10-11}
\textbf{Method} & \textbf{AUC} & Fall. & Exact & Approx & Packed & Fall. & CT-PT & CT-CT & CT-PT & CT-CT & \textbf{Time (s)} \\
\midrule
Non-SIMD (Baseline) & 0.7973 & 10000 & 0 & 0 & 0 & 1021247 & 0 & 0 & 4006774 & 39107 & 7751.7 \\
TGHE-Base ($\tau{=}0.50$) & 0.7950 & 104 & 6434 & 3462 & 10384 & 25054 & 38496 & 368 & 98028 & 1094 & 576.0 \\
TGHE-Collapse & 0.7959 & 0 & 10000 & 0 & 10252 & 0 & 38330 & 379 & 0 & 0 & 115.9 \\
\bottomrule
\end{tabular}
\vspace{2pt}
\par\noindent\footnotesize\textit{P.}\ = packed (SIMD) path; \textit{F.}\ = fallback (sequential) path. CT-PT and CT-CT denote ciphertext-plaintext and ciphertext-ciphertext multiplications, respectively.
\end{table*}

\subsection{Predictive Accuracy}
\label{sec:exp_utility}

Table~\ref{tab:utility} reports test ROC-AUC. Within the encrypted block, TGHE's packing and optimization contribute minimal additional loss: TGHE-Collapse achieves 0.7959 vs.\ 0.7973 for the Non-SIMD baseline (0.0014 difference, less than 0.2\% relative). Topology Collapse preserves higher accuracy than Approximate Template Fitting (0.7959 vs.\ 0.7950) because its pre-aggregation is algebraically exact. The gap between the plaintext GEARSage backbone (0.8460) and the encrypted block (0.7973) reflects polynomial activation approximation and degree capping, adaptations required by all HE-based GNN approaches, not by TGHE's packing stage.

\subsection{End-to-End Inference Performance}
\label{sec:exp_e2e}

Table~\ref{tab:e2e} reports end-to-end latency and cryptographic operation counts for the full 10{,}000-query workload.

\subsubsection{Non-SIMD (Baseline)}
Sequential execution processes each query independently, consuming 1{,}021{,}247 rotations and over 4M CT-PT operations. Total runtime exceeds 2 hours (7{,}751.7\,s) for 10{,}000 queries.

\subsubsection{TGHE-Base (13.5$\times$ speedup)}
Template-batched SIMD routes 98.96\% of queries through the packed path (6{,}434 exact + 3{,}462 approximate matches), reducing runtime to 576.0\,s. However, the 104 remaining fallback queries still account for 25{,}054 rotations (over twice the packed-path total of 10{,}384), confirming the Amdahl's Law bottleneck of Section~\ref{sec:longtail_bottleneck}.

\subsubsection{TGHE-Collapse (66.9$\times$ speedup)}
Topology Collapse eliminates the fallback path entirely. Every query achieves exact SIMD matching through degree-group templates. Runtime drops to 115.9\,s with only 10{,}252 rotations (a 99.0\% reduction) and 379 CT-CT operations, compared to 39{,}107 under the Non-SIMD baseline. This represents a 66.9$\times$ speedup over Non-SIMD and 5.0$\times$ over TGHE-Base. The threshold $\tau$ in Approximate Template Fitting provides a smooth accuracy-throughput dial: relaxing from $\tau{=}1.00$ to $\tau{=}0.50$ reduces runtime by 4.3$\times$ with only 0.0013 AUC loss, with accuracy flooring at 0.7948 below $\tau{=}0.25$.

\FloatBarrier

\subsection{Topology Leakage Profile}
\label{sec:exp_leakage}

TGHE adopts a leakage-aware design: rather than claiming full topology hiding, we explicitly characterize all information visible to the server. Under \textbf{TGHE-Base}, the server observes template signatures and surrogate padding volumes, enabling reconstruction of the tree-level ego-graph structure, but not the original subgraph, as treeification removes cycles and shared-node information. Under \textbf{TGHE-Collapse}, hop-2 node identities are hidden by client-side pre-aggregation; the server learns only the hop-1 degree $d_{\mathrm{eff}}(r)$ and per-hop-1 child counts, which is strictly coarser. In both modes, edge/time correction terms require transmitting plaintext edge metadata (type, direction, timestamp), the most sensitive residual, revealing relationship categories and temporal patterns. A hardened deployment could eliminate this by having the client pre-compute correction terms locally. Feature vectors and model outputs remain protected by CKKS semantic security throughout.

\balance
\section{Conclusion}
\label{sec:conclusion}
We presented TGHE, a framework for privacy-preserving GNN inference that shifts from the transductive paradigm of prior HE-GNN systems to an ego-centric design suited to large-scale, dynamic financial graphs in edge-cloud deployments. By extracting and canonicalizing local ego-graphs at the edge, TGHE decouples per-query cost from the global graph, enabling encrypted inference at million-node scale. A central contribution is the identification and exploitation of a template phenomenon: local computation trees in transaction graphs converge into a small set of structural shapes, enabling template-based CKKS ciphertext packing for SIMD-parallel evaluation. Two long-tail optimizers (Approximate Template Fitting and Topology Collapse) maximize SIMD coverage, with Topology Collapse eliminating the fallback path entirely. Experiments on DGraphFin demonstrate substantial end-to-end speedup with negligible accuracy loss, validating TGHE's applicability to real-world financial graph inference at scale.

\section*{Acknowledgment}
This work was partially supported by the Australian Research Council (ARC) through Linkage Project LP240100523, Blockchain Based Quantum Safe for Secure Digital Medical Passport. The authors gratefully acknowledge the support of the project partners and collaborators.

\bibliographystyle{IEEEtran}
\bibliography{references}

@article{corso2024graph,
  author    = {Corso, Gabriele and Stark, Hannes and Jegelka, Stefanie and Jaakkola, Tommi and Barzilay, Regina},
  title     = {Graph Neural Networks},
  journal   = {Nature Reviews Methods Primers},
  volume    = {4},
  number    = {1},
  pages     = {17},
  year      = {2024},
}

@article{acar2018survey,
  author    = {Acar, Abbas and Aksu, Hidayet and Uluagac, A. Selcuk and Conti, Mauro},
  title     = {A Survey on Homomorphic Encryption Schemes: Theory and Implementation},
  journal   = {ACM Computing Surveys},
  volume    = {51},
  number    = {4},
  pages     = {1--35},
  year      = {2018},
}

@inproceedings{ran2022cryptogcn,
  author    = {Ran, Ran and Xu, Nuo and Wang, Wei and Quan, Gang and Yin, Jianming and Wen, Wujie},
  title     = {{CryptoGCN}: Fast and Scalable Homomorphically Encrypted Graph Convolutional Network Inference},
  booktitle = {Advances in Neural Information Processing Systems (NeurIPS)},
  volume    = {35},
  year      = {2022},
  url       = {https://openreview.net/forum?id=VeQBBm1MmTZ},
}

@inproceedings{peng2023lingcn,
  author    = {Peng, Hongwu and Ran, Ran and Luo, Yukui and Zhao, Jiahui and Huang, Shaoyi and Thorat, Kishor and Geng, Tong and Wang, Chenghong and Xu, Xiaolin and Wen, Wujie and Ding, Caiwen},
  title     = {{LinGCN}: Structural Linearized Graph Convolutional Network for Homomorphically Encrypted Inference},
  booktitle = {Advances in Neural Information Processing Systems (NeurIPS)},
  year      = {2023},
  url       = {https://openreview.net/forum?id=5loV5tVzsY},
}

@inproceedings{ran2023penguin,
  author    = {Ran, Ran and Xu, Nuo and Liu, Tao and Wang, Wei and Quan, Gang and Wen, Wujie},
  title     = {{Penguin}: Parallel-Packed Homomorphic Encryption for Fast Graph Convolutional Network Inference},
  booktitle = {Advances in Neural Information Processing Systems (NeurIPS)},
  year      = {2023},
  url       = {https://proceedings.neurips.cc/paper_files/paper/2023/hash/3cc685788a311fa35d8d41df93e288ca-Abstract-Conference.html},
}

@inproceedings{ficgcn2024,
  author    = {Kan, Zhehan and Han, Haixin and Shi, Shuohan and Hua, Tianxiang and Lu, Haibin and Li, Xiang and Mu, Jian and Hu, Xiaobing},
  title     = {{FicGCN}: Unveiling the Homomorphic Encryption Efficiency from Irregular Graph Convolutional Networks},
  booktitle = {Proceedings of the 42nd International Conference on Machine Learning (ICML)},
  series    = {Proceedings of Machine Learning Research},
  volume    = {267},
  pages     = {28832--28848},
  publisher = {PMLR},
  year      = {2025},
  url       = {https://proceedings.mlr.press/v267/kan25a.html},
}

@inproceedings{cheon2017ckks,
  author    = {Cheon, Jung Hee and Kim, Andrey and Kim, Miran and Song, Yongsoo},
  title     = {Homomorphic Encryption for Arithmetic of Approximate Numbers},
  booktitle = {Advances in Cryptology -- ASIACRYPT 2017},
  series    = {Lecture Notes in Computer Science},
  volume    = {10624},
  pages     = {409--437},
  publisher = {Springer},
  year      = {2017},
  url       = {https://link.springer.com/chapter/10.1007/978-3-319-70694-8_15},
}

@misc{pyg2025dgraphfin,
  author       = {{PyTorch Geometric Team}},
  title        = {torch\_geometric.datasets.{DGraphFin} documentation},
  year         = {2025},
  note         = {Accessed via PyG documentation; includes dataset statistics},
  url          = {https://pytorch-geometric.readthedocs.io/en/2.7.0/generated/torch_geometric.datasets.DGraphFin.html},
}

@inproceedings{huang2022dgraph,
  author    = {Huang, Xuanwen and Yang, Yang and Wang, Yang and Wang, Chunping and Zhang, Zhisheng and Xu, Jiarong and Chen, Lei and Vazirgiannis, Michalis},
  title     = {{DGraph}: A Large-Scale Financial Dataset for Graph Anomaly Detection},
  booktitle = {Advances in Neural Information Processing Systems (NeurIPS), Datasets and Benchmarks Track},
  year      = {2022},
  url       = {https://proceedings.neurips.cc/paper_files/paper/2022/hash/8f1918f71972789db39ec0d85bb31110-Abstract-Datasets_and_Benchmarks.html},
}

@inproceedings{he2021stealing,
  author    = {He, Xinlei and Jia, Jinyuan and Backes, Michael and Gong, Neil Zhenqiang and Zhang, Yang},
  title     = {Stealing Links from Graph Neural Networks},
  booktitle = {30th USENIX Security Symposium (USENIX Security 21)},
  pages     = {2669--2686},
  year      = {2021},
  url       = {https://www.usenix.org/system/files/sec21summer_he.pdf},
}

@inproceedings{fu2025safeguarding,
  author    = {Fu, Jiaqi and Hong, Yuan and Chen, Zixin and Wang, Wendy Hui},
  title     = {Safeguarding Graph Neural Networks against Topology Inference Attacks},
  booktitle = {Proceedings of the 32nd ACM Conference on Computer and Communications Security (ACM CCS)},
  year      = {2025},
  note      = {Also available as arXiv:2509.05429},
  url       = {https://arxiv.org/abs/2509.05429},
}

@misc{sealcrypto2024,
  author       = {{Microsoft Research}},
  title        = {{Microsoft SEAL} (release 4.1)},
  year         = {2024},
  url          = {https://github.com/microsoft/SEAL},
  note         = {Homomorphic encryption library},
}

@misc{gearsage2022,
  author       = {Li, Jintang and Yu, Zhouxin and Sun, Wangbin and Jin, Xinzhou and Wang, Qichao and Chen, Liang},
  title        = {{GEARSage}: Graph Edge-Aware {GraphSAGE} for {DGraphFin}},
  year         = {2022},
  howpublished = {7th Finvolution Data Science Competition, technical report},
  url          = {https://github.com/storyandwine/GEARSage-DGraphFin},
  note         = {{Sun Yat-sen University}, SYSU-GEAR team}
}

@inproceedings{kipf2017gcn,
  author    = {T. N. Kipf and M. Welling},
  title     = {Semi-supervised classification with graph convolutional networks},
  booktitle = {Proc.\ International Conference on Learning Representations (ICLR)},
  year      = {2017}
}

@inproceedings{hamilton2017graphsage,
  author    = {W. Hamilton and Z. Ying and J. Leskovec},
  title     = {Inductive representation learning on large graphs},
  booktitle = {Advances in Neural Information Processing Systems (NeurIPS)},
  volume    = {30},
  year      = {2017}
}

@inproceedings{velickovic2018gat,
  author    = {P. Veli{\v{c}}kovi{\'{c}} and G. Cucurull and A. Casanova and A. Romero and P. Li{\`{o}} and Y. Bengio},
  title     = {Graph attention networks},
  booktitle = {Proc.\ International Conference on Learning Representations (ICLR)},
  year      = {2018}
}

@article{frasca2020sign,
  author  = {F. Frasca and E. Rossi and D. Eynard and B. Chamberlain and M. Bronstein and F. Monti},
  title   = {{SIGN}: Scalable inception graph neural networks},
  journal = {arXiv preprint arXiv:2004.11198},
  year    = {2020}
}

@inproceedings{yun2019graph_transformer,
  author    = {S. Yun and M. Jeong and R. Kim and J. Kang and H. J. Kim},
  title     = {Graph transformer networks},
  booktitle = {Advances in Neural Information Processing Systems (NeurIPS)},
  volume    = {32},
  year      = {2019}
}

@inproceedings{halevi2018faster,
  author    = {Shai Halevi and Victor Shoup},
  title     = {Faster Homomorphic Linear Transformations in {HElib}},
  booktitle = {Advances in Cryptology -- {CRYPTO} 2018},
  series    = {Lecture Notes in Computer Science},
  volume    = {10991},
  pages     = {93--120},
  publisher = {Springer},
  year      = {2018}
}

@inproceedings{weber2019elliptic,
  author    = {Mark Weber and Giacomo Domeniconi and Jie Chen and Daniel Karl I. Weidele and Claudio Bellei and Tom Robinson and Charles E. Leiserson},
  title     = {Anti-Money Laundering in Bitcoin: Experimenting with Graph Convolutional Networks for Financial Forensics},
  booktitle = {KDD Workshop on Anomaly Detection in Finance},
  year      = {2019},
  note      = {arXiv:1908.02591}
}

@misc{suzumura2021amlsim,
  author       = {Toyotaro Suzumura and Hiroki Kanezashi},
  title        = {{AMLSim}: A Multi-Agent Simulator of Anti-Money Laundering},
  howpublished = {\url{https://github.com/IBM/AMLSim}},
  year         = {2021}
}

@inproceedings{altman2023ibmaml,
  author    = {Erik Altman and Jovan Blanu\v{s}a and Luc von Niederh\"{a}usern and B\'{e}ni Egressy and Andreea Anghel and Kubilay Atasu},
  title     = {Realistic Synthetic Financial Transactions for Anti-Money Laundering Models},
  booktitle = {Advances in Neural Information Processing Systems ({NeurIPS}), Datasets and Benchmarks Track},
  year      = {2023}
}

\end{document}